\documentclass[12pt]{article}
\usepackage{graphicx}
\usepackage{lscape}
\usepackage{citesort}
\usepackage{amssymb}
\usepackage{appendix}
\usepackage{multirow}

\begin{document}
\def\qq{\langle \bar q q \rangle}
\def\uu{\langle \bar u u \rangle}
\def\dd{\langle \bar d d \rangle}
\def\sp{\langle \bar s s \rangle}
\def\GG{\langle g_s^2 G^2 \rangle}
\def\Tr{\mbox{Tr}}
\def\figt#1#2#3{
        \begin{figure}
        $\left. \right.$
        \vspace*{-2cm}
        \begin{center}
        \includegraphics[width=10cm]{#1}
        \end{center}
        \vspace*{-0.2cm}
        \caption{#3}
        \label{#2}
        \end{figure}
	}
	
\def\figb#1#2#3{
        \begin{figure}
        $\left. \right.$
        \vspace*{-1cm}
        \begin{center}
        \includegraphics[width=10cm]{#1}
        \end{center}
        \vspace*{-0.2cm}
        \caption{#3}
        \label{#2}
        \end{figure}
                }

\def\ds{\displaystyle}
\def\beq{\begin{equation}}
\def\eeq{\end{equation}}
\def\bea{\begin{eqnarray}}
\def\eea{\end{eqnarray}}
\def\beeq{\begin{eqnarray}}
\def\eeeq{\end{eqnarray}}
\def\ve{\vert}
\def\vel{\left|}
\def\ver{\right|}
\def\nnb{\nonumber}
\def\ga{\left(}
\def\dr{\right)}
\def\aga{\left\{}
\def\adr{\right\}}
\def\lla{\left<}
\def\rra{\right>}
\def\rar{\rightarrow}
\def\lrar{\leftrightarrow}  
\def\nnb{\nonumber}
\def\la{\langle}
\def\ra{\rangle}
\def\ba{\begin{array}}
\def\ea{\end{array}}
\def\tr{\mbox{Tr}}
\def\ssp{{\Sigma^{*+}}}
\def\sso{{\Sigma^{*0}}}
\def\ssm{{\Sigma^{*-}}}
\def\xis0{{\Xi^{*0}}}
\def\xism{{\Xi^{*-}}}
\def\qs{\la \bar s s \ra}
\def\qu{\la \bar u u \ra}
\def\qd{\la \bar d d \ra}
\def\qq{\la \bar q q \ra}
\def\gGgG{\la g^2 G^2 \ra}
\def\q{\gamma_5 \not\!q}
\def\x{\gamma_5 \not\!x}
\def\g5{\gamma_5}
\def\sb{S_Q^{cf}}
\def\sd{S_d^{be}}
\def\su{S_u^{ad}}
\def\sbp{{S}_Q^{'cf}}
\def\sdp{{S}_d^{'be}}
\def\sup{{S}_u^{'ad}}
\def\ssp{{S}_s^{'??}}

\def\sig{\sigma_{\mu \nu} \gamma_5 p^\mu q^\nu}
\def\fo{f_0(\frac{s_0}{M^2})}
\def\ffi{f_1(\frac{s_0}{M^2})}
\def\fii{f_2(\frac{s_0}{M^2})}
\def\O{{\cal O}}
\def\sl{{\Sigma^0 \Lambda}}
\def\es{\!\!\! &=& \!\!\!}
\def\ap{\!\!\! &\approx& \!\!\!}
\def\ar{&+& \!\!\!}
\def\ek{&-& \!\!\!}
\def\kek{\!\!\!&-& \!\!\!}
\def\cp{&\times& \!\!\!}
\def\se{\!\!\! &\simeq& \!\!\!}
\def\eqv{&\equiv& \!\!\!}
\def\kpm{&\pm& \!\!\!}
\def\kmp{&\mp& \!\!\!}
\def\mcdot{\!\cdot\!}
\def\erar{&\rightarrow&}


\def\simlt{\stackrel{<}{{}_\sim}}
\def\simgt{\stackrel{>}{{}_\sim}}

\def\olra{\stackrel{\leftrightarrow}}
\def\ola{\stackrel{\leftarrow}}
\def\ora{\stackrel{\rightarrow}}


\title{
         {\Large
                 {\bf
Radiative decays of the heavy tensor mesons in 
light cone QCD sum rules
                 }
         }
      }

\author{\vspace{1cm}\\
{\small T. M. Aliev\thanks {e-mail:
taliev@metu.edu.tr}\,\,,
M. Savc{\i}\thanks
{e-mail: savci@metu.edu.tr}} \\
{\small Physics Department, Middle East Technical University,
06531 Ankara, Turkey} }
\date{}

\begin{titlepage}
\maketitle
\thispagestyle{empty}

\begin{abstract}
The transition form factors of the radiative decays of the heavy tensor
mesons to heavy pseudoscalar and heavy vector mesons are calculated in
the framework of the light cone QCD sum rules method at the point $Q^2=0$.
Using the obtained values of the transition form factors at the point $Q^2=0$
the corresponding decay widths are estimated. The results show that
the radiative decays of the heavy--light tensor mesons could potentially 
be measured in the future planned experiments at LHCb.
\end{abstract}

\vspace{1cm}
\end{titlepage}

\section{Introduction}

With the recent developments in the experimental techniques, many new
particles have been discovered \cite{Resr01,Resr02,Resr03,Resr04,Resr05}.
Part of the newly discovered
particles have already been predicted by the quark model. But the rest is not
expected to be foreseen by the quark model, and understanding their
properties requires new perspective beyond the conventional quark model.
The heavy tensor mesons ${\cal D}_2(2460)$, ${\cal D}_{s_2}(2573)$,
$B_2(5747)$ and $B_{s_2}(5840)$ predicted by the
conventional quark model have all been discovered in the experiments, and their
masses and decay widths have been measured \cite{Resr06}. More refined analysis in
studying the properties of these particles will be conducted at LHCb and
BELLE--2.

Following the discovery of the heavy tensor mesons their strong,
electromagnetic and weak decays need to be investigated. Note that the
strong decays ${\cal D}_2^0(2460) \to {\cal D}^+({\cal D}^{\ast +}) \pi^-$,
${\cal D}_2^+(2460) \to {\cal D}^0 \pi^+$
\cite{Resr07,Resr08,Resr09,Resr10},
${\cal D}_{s_2}^+(2573) \to {\cal D}^0 K^+$ \cite{Resr07},
$B_2^0(5747) \to B^{\ast +} \pi^-$ \cite{Resr11,Resr12} and
$B_{s_2}^0(5840) \to B^+ K^-$\cite{Resr11,Resr12}
have already been observed in the experiments.

These observations have stimulated a chain of many studies. For
example, the strong coupling constants of the aforementioned decays have
been calculated in the framework of the three--point sum rules
\cite{Resr13,Resr14,Resr15}, and in the light cone QCD sum rules (LCSR)
methods \cite{Resr16}.

In the present work, we study the radiative decays of the heavy tensor mesons
in the framework of the LCSR. Radiative decays constitute one of the most promising
classes of decays in gathering information about the electromagnetic
properties, which are important to reveal the internal structure of the
hadrons. It should be emphasized
here that, so far the radiative decays of the heavy tensor mesons have not
yet been observed in the experiments, and our results might indicate that
these decays can potentially be measured at LHCb.

The paper is organized as follows: In section 2, we formulate the LCSR
for the transition form factors at the point $Q^2=0$. In section 3,
we perform numerical analysis of these form factors at the point $Q^2=0$
and calculate the corresponding decay widths. The last section contains our
conclusion.

\section{Light cone QCD sum rules for the heavy tensor $\to$ heavy
pseudoscalar(vector) meson $+$ photon}

Before presenting the details of the calculation,
few words about our notation are in order.
In the present work the states of the heavy tensor, heavy vector, and heavy pseudoscalar
mesons are denoted by the generic notations $T_Q$, $V_Q$, and $P_Q$, respectively.
 
The $T_Q \to P_Q(V_Q) \gamma$ decay is described by the following
correlator:
\bea
\label{eesr01}
\Pi_{\mu\nu\alpha(\rho)} (p,q) = - \int d^4x \int d^4y e^{i(px+qy)}
\lla 0 \vel J_{{T_Q}\mu\nu}(x) J_{\alpha}^{e\ell}(y) J_{P_Q}(0)
\left(J_{{V_Q}\rho}(0)\right) \ver
0 \rra~,
\eea
where 
\bea
\label{nolabel01}
J_{{T_Q}\mu\nu}(x) \es {1 \over 2} \left[\bar{q}(x) \gamma_\mu 
\olra{\cal D}_\nu\!(x)Q(x) + \bar{q}(x) \gamma_\nu 
\olra{\cal D}_\mu\!(x)Q(x)\right]~,\nnb \\
J_{P_Q} (J_{{V_Q}\rho}) \es \bar{q} i\gamma_5 Q(\bar{q} \gamma_\rho Q)\nnb
\eea
are the interpolating currents of the heavy tensor, heavy pseudoscalar
(heavy vector) mesons, respectively, and
\bea
\label{nolabel02}
J_{\alpha}^{e\ell}(y) = e_q \bar{q} \gamma_\alpha q + e_Q \bar{Q}
\gamma_\alpha Q~,\nnb
\eea
is the electromagnetic current, where $e_q$ and $e_Q$ are the electric
charges of the light and heavy quarks, respectively.   
The covariant derivative $\olra{\cal D}_\mu\!(x)$ is defined as,
\bea
\label{nolabel03}
\olra{\cal D}_\mu\!(x) = {1\over 2} \Big[
\ora{\cal D}_\mu\!(x) -
\ola{\cal D}_\mu\!(x) \Big]\nnb
\eea
where
\bea
\label{nolabel04}
\ora{\cal D}_\mu\!(x) \es \ora{\partial}_\mu\!(x) - i
g{\lambda^a\over 2} A_\mu^a (x) ~, \nnb \\
\ola{\cal D}_\mu\!(x) \es \ola{\partial}_\mu\!(x) + i
g {\lambda^a\over 2} A_\mu^a (x) ~.\nnb
\eea
In this expression  $\lambda^a$ are the Gell--Mann matrices, $A_\mu^a (x)$ is the external
field.

The correlator given in Eq. (\ref{eesr01}) can be rewritten in the presence
of the electromagnetic background field of a plane wave
\bea
\label{nolabel05}
F_{\mu\nu} = i (\varepsilon_\mu^{(\gamma)} q_\nu - 
\varepsilon_\nu^{(\gamma)} q_\mu~)\nnb
\eea
in the following form,
\bea
\label{eesr02}
\Pi_{\mu\nu\alpha(\rho)} \varepsilon^{(\gamma)\alpha} = 
i \int d^4x e^{ipx} \lla 0 \vel J_{{T_Q}\mu\nu}(x) J_{P_Q}
(J_{{V_Q}\rho})(0) \ver 0 \rra_F
\eea
where the subscript $F$ stands for the vacuum expectation value evaluated
in the presence of the background  electromagnetic field $F_{\mu\nu}$. The
expression of the correlation function given in Eq. (\ref{eesr01}) can be
obtained by expanding Eq. (\ref{eesr02}) in powers of the background field 
by taking into account only the terms linear in $F_{\mu\nu}$ which corresponds  
to the single photon emission (for more details about the background field
method and its applications, see \cite{Resr17,Resr18}).

In order to calculate any physical quantity in framework of the QCD sum
rules method, the correlation function needs to be computed in two different
kinematical domains. On the one side, the main contribution to the
correlation function (\ref{eesr02}) originates in the region where
$p^2\simeq m_{T_Q}^2$ and
$(p+q)^2\simeq m_{P_Q}^2(m_{V_Q}^2)$. On the other side the same correlation
function can be investigated in the deep Euclidean domain where $p^2\ll0$,
$(p+q)^2\ll0$, using the operator product expansion (OPE). As is well known,
in the LCSR method the OPE is performed over the twists of the operators rather
than their canonical dimensions, which is the case in the standard sum rules
approach. The physical part of the correlation function Eq. (\ref{eesr01}) is
obtained by inserting a complete set of the corresponding mesonic states,
and then isolating the ground state tensor and pseudoscalar (vector) mesons,
as shown below,
\bea
\label{eesr03}
\Pi_{\mu\nu\alpha(\rho)}(p,q) = {\lla 0 \vel J_{{T_Q}\mu\nu} \ver T_Q(p) \rra
\over (m_{T_Q}^2 -p^2)}{\lla T_Q(p) \vel J_\alpha^{e\ell} \ver P_Q(V_Q)(p+q) \rra
\over [m_{P_Q(V_Q)}^2 -(p+q)^2]}\lla P_Q(V_Q) \vel J_{P_Q}
(J_{{V_Q}\rho}) \ver 0 \rra + \cdots~,
\eea
where dots denote 	the higher state contributions, and $p^\prime=p+q$.
The matrix elements in Eq. (\ref{eesr03}) are defined as follow:
\bea
\label{eesr04}
\lla 0 \ver J_{{T_Q}\mu\nu} \vel T_Q(p)\rra \es f_{T_Q} m_{T_Q}^3
\epsilon_{\mu\nu}(p)~,\nnb \\
\lla P_Q \vel J_{P_Q} \ver 0 \rra \es {f_{P_Q} m_{P_Q}^2 \over
m_Q+m_q}~,\nnb \\
\lla T_Q(p) \vel J_\alpha^{e\ell} \ver P_Q(p+q) \rra \es g
\varepsilon_{\alpha\rho\lambda\tau} \epsilon^{\rho\xi}p_\xi^\prime p^\lambda
q^\tau~,\nnb\\
\lla V_Q(\varepsilon,p^\prime) \vel J_{{V_Q}\rho} \ver 0 \rra \es f_{V_Q} m_{V_Q}
\varepsilon_\rho^\ast~.\nnb \\
\lla T_Q(p) \vel J_\alpha^{e\ell} \ver V_Q(p^\prime) \rra \es
h_1 \epsilon_{\alpha\beta} \varepsilon^\beta +
h_2 \epsilon_{\alpha\beta} p^{\prime \beta} (\varepsilon\mcdot p)+
h_3 \varepsilon_\alpha \epsilon_{\beta\tau}
p^{\prime\beta}p^{\prime\tau} +
h_4^\prime p_\alpha \epsilon_{\beta\tau}\varepsilon^\beta p^{\prime \tau}\nnb \\
\ar h_5^\prime p_\alpha^\prime \epsilon_{\beta\tau} \varepsilon^\beta
p^{\prime\tau} +
h_6^\prime p_\alpha \epsilon_{\beta\tau} p^{\prime\beta}
p^{\prime\tau}(\varepsilon\mcdot p) +
h_7^\prime p_\alpha^\prime \epsilon_{\beta\tau} p^{\prime\beta} p^{\prime\tau}
(\varepsilon\mcdot p)~.
\eea
In these expressions $\epsilon_{\alpha\beta}$ and $\varepsilon_\alpha$, are the
tensor and vector meson polarizations, $f_{P_Q}$ and $f_{V_Q}$
are the decay constants of the heavy pseudoscalar and vector mesons,
$m_{P_Q}$ and $m_{V_Q}$ are their masses, $m_Q$ and $m_q$ are the heavy
and light quark masses,
$g$ and $h_i$ are the form factors responsible
for the $T_Q \to P_Q$ and $T_Q \to V_Q$ transitions, respectively.  

Substituting these matrix elements into the physical part of the
corresponding correlation functions given in
Eq. (\ref{eesr01}), we get
\bea
\label{eesr05}
\Pi_{\mu\nu\alpha}\varepsilon^{(\gamma)\alpha}(q) \es {1\over m_{T_Q}^2-p^2}
{1\over m_{P_Q}^2 - p^{\prime 2}} f_{T_Q} m_{T_Q}^3 \epsilon_{\mu\nu} {f_{P_Q}
m_{P_Q}^2 \over
m_Q+m_q} g \varepsilon_{\alpha\rho\lambda\tau} p^\lambda q^\tau
p^{\prime}_\xi \epsilon^{\rho\xi} \varepsilon^{(\gamma)\alpha}(q) \\
\label{eesr06}
\Pi_{\mu\nu\alpha\rho}\varepsilon^{(\gamma)\alpha}(q) \es
{1\over m_{T_Q}^2-p^2} {1\over m_{V_Q}^2 - p^{\prime 2}} f_{T_Q} m_{T_Q}^3
f_{V_Q} m_{V_Q}
\epsilon_{\mu\nu}(p)\varepsilon^{(\gamma)\alpha}\varepsilon_\rho^\ast \Big\{
h_1 \epsilon_{\alpha\beta} \varepsilon^\beta +
h_2 \epsilon_{\alpha\beta} p^{\prime \beta} (\varepsilon\mcdot p) \nnb \\
\ar h_3 \varepsilon^\alpha \epsilon_{\beta\tau}
p^{\prime\beta}p^{\prime\tau} +
h_4^\prime p_\alpha \epsilon_{\beta\tau}\varepsilon^\beta p^{\prime \tau}\nnb
+ h_5^\prime p_\alpha^\prime \epsilon_{\beta\tau} \varepsilon^\beta
p^{\prime\tau} + 
h_6^\prime p_\alpha \epsilon_{\beta\tau} p^{\prime\beta}
p^{\prime\tau} (\varepsilon\mcdot p) \nnb \\
\ar h_7^\prime p_\alpha^\prime \epsilon_{\beta\tau} p^{\prime\beta} p^{\prime\tau}
(\varepsilon\mcdot p)\Big\}~.
\eea
Performing summation over the spins of the the tensor and vector
mesons with the help of the identities,
\bea
\label{eesr07}
\epsilon_{\mu\nu}(p) \epsilon_{\alpha\beta}^\ast(p) \es
{1\over 2} {\cal P}_{\mu\alpha} {\cal P}_{\nu\beta} +
{1\over 2} {\cal P}_{\mu\beta} {\cal P}_{\nu\alpha} -
{1\over 3} {\cal P}_{\mu\nu} {\cal P}_{\alpha\beta}~,
~\mbox{where~}
{\cal P}_{\alpha\beta} = -g_{\alpha\beta} + {p_\alpha  
p_\beta \over m_{T_Q}^2}~,\nnb\\ 
\varepsilon_\alpha(p^\prime) \varepsilon_\beta^\ast(p^\prime) \es
{\cal P}_{\alpha\beta}^\prime~,~\mbox{where~}
{\cal P}_{\alpha\beta}^\prime = -g_{\alpha\beta} + {p_\alpha^\prime
p_\beta^\prime \over m_{V_Q}^2}~,
\eea
for the physical parts of the correlation functions we have,
\bea
\label{eesr08}
\Pi_{\mu\nu\alpha}\varepsilon^{(\gamma)\alpha}(q) \es
{f_{T_Q} m_{T_Q}^3\over m_{T_Q}^2-p^2}
{f_{P_Q} m_{P_Q}^2\over m_{P_Q}^2 - p^{\prime 2}} 
{\varepsilon^{(\gamma)\alpha}(q)\over m_Q+m_q} g
\Bigg\{{1\over 2} \varepsilon_{\alpha\mu\lambda\tau} p^{\lambda}
q^\tau \Bigg(p_\nu^\prime -{p_\nu (p\mcdot p^\prime) \over
m_{T_Q}^2} \Bigg) \nnb \\
\ar (\mu \lrar \nu) \Bigg\} + \cdots~,\\
\label{eesr09}
\Pi_{\mu\nu\alpha\rho}\varepsilon^{(\gamma)\alpha}(q) \es
{f_{T_Q} m_{T_Q}^3\over m_{T_Q}^2-p^2}
{f_{V_Q} m_{V_Q}\over m_{P_Q}^2 - p^{\prime 2}} \varepsilon^{(\gamma)\alpha}(q)
\Bigg\{ {1\over 2}\Bigg({\cal P}_{\mu\alpha} {\cal P}_{\nu\beta} +
{\cal P}_{\mu\beta} {\cal P}_{\nu\alpha} - {2\over 3} {\cal P}_{\mu\nu}
{\cal P}_{\alpha\beta}\Bigg)\nnb \\
\cp \Big[h_1 {\cal P}_\rho^{\prime\beta} +
h_2 q^\beta  p^\xi {\cal P}_{\rho\xi}^\prime\Big] +
{1\over 2}\Bigg({\cal P}_{\mu\lambda} {\cal P}_{\nu\tau} +
{\cal P}_{\mu\tau} {\cal P}_{\nu\lambda} - {2\over 3} {\cal P}_{\mu\nu}     
{\cal P}_{\lambda\tau}\Bigg) \nnb \\
\cp \Big[h_3 q^\lambda q^\tau 
{\cal P}_{\alpha\rho}^\prime +h_4 p_\alpha q^\tau {\cal P}_\rho^{\prime\lambda} +
h_5 p_\alpha  q^\lambda q^\tau p^\xi {\cal P}_{\rho\xi}^\prime\Big]+ (\mu
\lrar \nu) \Bigg\} + \cdots~,
\eea
where $h_4=h_4^\prime + h_5^\prime$, and $h_5=h_6^\prime + h_7^\prime$;
and dots denote the contributions coming from the excited states and
continuum. 

In order to determine the form factor $g$ for the $T_Q \to P_Q \gamma$
transition we choose the coefficient
of the structure $\varepsilon_{\alpha\mu\lambda\tau}p^\lambda q^\tau q^\nu$.
But for the vector $T_Q \to V_Q \gamma$ transition the situation is much
more complicated, for which there are numerous structures.
In this case not all transition form factors are independent. Indeed using
the gauge invariance one can easily obtain,
\bea
\label{nolabel10}
h_1 + h_4 (p\mcdot q) \es 0 \nnb \\
-h_2 + h_3 -h_5 (p\mcdot q)\es 0~. \nnb
\eea
It follows from these relations that we have only three independent form
factors. Using these relations the matrix element $\Pi_{\mu\nu\alpha\rho}
\varepsilon^{(\gamma)\alpha}$ can be written as,
\bea
\label{eesr10}
\Pi_{\mu\nu\alpha\rho}\varepsilon^{(\gamma)\alpha} (q) \es
{f_{T_Q} m_{T_Q}^3\over m_{T_Q}^2-p^2}
{f_{V_Q} m_{V_Q}\over m_{P_Q}^2 - p^{\prime 2}}
\Bigg\{{h_1\over 2}\Big[\varepsilon^{(\gamma)\alpha} -{1\over p\mcdot q}
(\varepsilon^{(\gamma)}\mcdot p) q^\alpha\Big] {\cal P}_\rho^{\prime\beta}
\Bigg({\cal P}_{\alpha\mu} {\cal P}_{\nu\beta} +
{\cal P}_{\mu\beta} {\cal P}_{\alpha\nu} \nnb \\
\ek {2\over 3} {\cal P}_{\mu\nu}
{\cal P}_{\alpha\beta}\Bigg) + {h_2\over 2}
\Bigg[\varepsilon^{(\gamma)\alpha}-
{1\over p\mcdot q}(\varepsilon^{(\gamma)}\mcdot p) q^\alpha\Big] 
{\cal P}_{\rho\xi}^\prime q^\beta q^\xi \Bigg({\cal
P}_{\alpha\mu} {\cal P}_{\nu\beta} +          
{\cal P}_{\mu\beta} {\cal P}_{\alpha\nu}
- {2\over 3} {\cal P}_{\mu\nu}  
{\cal P}_{\alpha\beta}\Bigg) \nnb \\
\ar {h_3\over 2} \Bigg[\varepsilon^{(\gamma)\alpha}-
{1\over p\mcdot q}(\varepsilon^{(\gamma)}\mcdot p) q^\alpha\Big] 
{\cal P}_{\alpha\rho}^\prime q^\beta q^\tau \Bigg(
{\cal P}_{\mu\beta} {\cal P}_{\nu\tau} +
{\cal P}_{\mu\tau} {\cal P}_{\nu\beta}
- {2\over 3} {\cal P}_{\mu\nu}
{\cal P}_{\beta\tau}\Bigg)~. 
\eea
As a result, in this
transition, we have three independent form factors $h_i,~(i=1,..,3)$, and hence
we need three independent equations to determine them. In other words,
three different structures are needed. In principle, any three structures
can be chosen in determining the three transition form factors. It is known
that the structures having the highest number of momenta makes the vacuum
expectation values of the higher dimensionality operators numerically less
important and the operator product expansion exhibits good convergence (see
for example \cite{Reesr19}). 
For this reason we choose the structures 
$(\varepsilon^{(\gamma)} \mcdot p)q _\mu g_{\nu\rho}$,
$(\varepsilon^{(\gamma)} \mcdot p) p_\mu q_\nu q_\rho$, and
$\varepsilon_\rho^{(\gamma)} q_\mu q_\nu$
in determining the form factors.

Having obtained the representation of the correlator function from the
physical side, our next job is to calculate it in the deep--Euclidean
domain using OPE. For this purpose, the explicit expressions of the interpolating
currents for the heavy tensor and pseudoscalar (vector) mesons should be
inserted into Eq. (\ref{eesr02}), as a result of which we get,
\bea
\label{eesr11}
\Pi_{\mu\nu\alpha(\rho)}\varepsilon^{(\gamma)\alpha}(q) =
i \int d^4x e^{ipx} \lla 0 \vel \Bigg[ {1\over 2} \bar{q}(x) \gamma_\mu 
\olra{\cal D}_\nu Q(x) + \mu \lrar \nu \Bigg]
\bar{Q}(0) i \gamma_5 q(0) \left(\bar{Q}(0) \gamma_\rho q(0) \right)\ver 0 \rra_F~.
\eea
In order to perform OPE, we need the expressions of the light and heavy
quark propagators in the presence of the gluonic and electromagnetic
background fields. In the Fock--Schwinger gauge, where the path ordering
exponents can be omitted, these propagators can be written as,
\bea
\label{eesr12}
S_q(x) \es {i\not\!{x} \over 2 \pi^2 x^4} - {im_q\over 4 \pi^2 x^2}
-{i\over 16 \pi^2 x^2} \int_0^1
du \Bigg\{g \Big[\bar{u} \rlap/{x} \sigma_{\alpha\beta} +
u\sigma_{\alpha\beta} \rlap/{x} \Big] G^{\alpha\beta} \nnb \\
\ar e_q\Big[\bar{u}\rlap/{x} \sigma_{\alpha\beta} + u \sigma_{\alpha\beta}
\rlap/{x} \Big]F^{\alpha\beta} (ux) \Bigg\}
- {im_q\over 32\pi^2} \int_0^1 \Big[g_s G_{\alpha\beta}
\sigma^{\alpha\beta} + e_q F_{\alpha\beta}G^{\alpha\beta}\Big] \nnb \\
\cp \Bigg(\ln {-x^2\Lambda^2\over 4} + 2 \gamma_E\Bigg)~,\\ \nnb \\
\label{eesr13}
S_Q(x) \es {m_Q^2 \over 4 \pi^2} \Bigg\{ {K_1(m_Q\sqrt{-x^2}) \over
\sqrt{-x^2}} + {i\rlap/{x} \over \left(\sqrt{-x^2}\right)^2}
K_2(m_Q\sqrt{-x^2}) \Bigg\} \nnb \\
\ek {g_s \over 16 \pi^2} \int_0^1 du
G_{\mu\nu}(ux) \left[ \left(\sigma^{\mu\nu} \rlap/x + \rlap/x
\sigma^{\mu\nu}\right) {K_1 (m_Q\sqrt{-x^2})\over \sqrt{-x^2}} +
2 \sigma^{\mu\nu} K_0(m_Q\sqrt{-x^2})\right]~,
\eea
where $K_i(m_Q\sqrt{-x^2})$ are the modified Bessel functions, $\Lambda$ is
the parameter separating the non--perturbative and perturbative domains,
whose value is calculated in \cite{Reesr20} to be $\Lambda = (0.5 \pm
0.1)~GeV$.
Note that the contributions of the nonlocal operators $\bar{q}G^2 q$,
$\bar{q}q\bar{q}q$ are small (see \cite{Resr19}), 
and these contributions
are all neglected in the Eqs. (\ref{eesr12}) and (\ref{eesr13}).

Using the explicit expressions of the heavy and light quark propagators the
correlator function(s) given in Eq. (\ref{eesr11}) can be calculated. The
correlator functions contain perturbative and nonperturbative parts. The
perturbative part corresponds to the case when a photon interacts with the
quark propagator perturbatively. The perturbative contribution is obtained
by taking the first two terms in the quark propagator into account, and a
photon field that interacts with the quark field perturbatively.

The non--perturbative contribution is obtained by replacing the light quark
propagator by
\bea
\label{nolabel06}
S_{\alpha\beta}(x-y) \rar -{1\over 4}
\left(\Gamma_k\right)_{\alpha\beta} \bar{q}^a \Gamma_k q^b~,\nnb
\eea
where $\Gamma_k = \Big\{ I,~\gamma_\mu,~\gamma_5,~i
\gamma_5\gamma_\mu,~\sigma_{\mu\nu}/\sqrt{2} \Big\}$ are the full set of
Dirac matrices. In this case there appear the matrix elements of two, three
particle non--local operators between the vacuum and the photon states. The
matrix elements are parametrized in terms of the photon distribution
amplitudes (DAs) as follows \cite{Resr17},
\bea
\label{nolabel44}
&&\langle \gamma(q) \vert  \bar q(x) \sigma_{\mu \nu} q(0) \vert  0
\rangle  = -i e_q \qq (\varepsilon_\mu^{(\gamma)} q_\nu - 
\varepsilon_\nu^{(\gamma)}
q_\mu) \int_0^1 du e^{i \bar{u} (q\cdot x)} \left(\chi \varphi_\gamma(u) +
\frac{x^2}{16} \mathbb{A}  (u) \right) \nnb \\ &&
-\frac{i}{2(q\mcdot x)}  e_q \qq \left[x_\nu \left(\varepsilon_\mu^{(\gamma)} - q_\mu
\frac{\varepsilon^{(\gamma)}\mcdot x}{q\mcdot x}\right) - x_\mu
- \left(\varepsilon_\nu^{(\gamma)} -
q_\nu \frac{\varepsilon^{(\gamma)}\mcdot x}{q\mcdot x}\right) \right] \int_0^1 du e^{i 
\bar{u} (q\cdot x)} h_\gamma(u)~,
\nnb \\
&&\langle \gamma(q) \vert  \bar q(x) \gamma_\mu q(0) \vert 0 \rangle
= e_q f_{3 \gamma} \left(\varepsilon_\mu^{(\gamma)} - q_\mu 
\frac{\varepsilon^{(\gamma)}\mcdot x}{q\mcdot x} \right) \int_0^1 du e^{i
\bar{u} (q\cdot x)} \psi^v(u)~,
\nnb \\
&&\langle \gamma(q) \vert \bar q(x) \gamma_\mu \gamma_5 q(0) \vert 0
\rangle  = - \frac{1}{4} e_q f_{3 \gamma} \epsilon_{\mu \nu \alpha
\beta } \varepsilon^{(\gamma)\nu} q^\alpha x^\beta \int_0^1 du e^{i 
\bar{u}( q \cdot x)} \psi^a(u)~,
\nnb \\
&&\langle \gamma(q) | \bar q(x) g_s G_{\mu \nu} (v x) q(0) \vert 0
\rangle = -i e_q \qq \left(\varepsilon_\mu^{(\gamma)} q_\nu -
\varepsilon_\nu^{(\gamma)}
q_\mu \right) \int {\cal D}\alpha_i e^{i (\alpha_{\bar q} + v
\alpha_g) (q\cdot x)} {\cal S}(\alpha_i)~,
\nnb \\
&&\langle \gamma(q) | \bar q(x) g_s \tilde G_{\mu \nu} i \gamma_5 (v
x) q(0) \vert 0 \rangle = -i e_q \qq \left(\varepsilon_\mu^{(\gamma)} q_\nu -
\varepsilon_\nu^{(\gamma)} q_\mu \right) \int {\cal D}\alpha_i e^{i
(\alpha_{\bar q} + v \alpha_g) (q\cdot x)} \tilde {\cal S}(\alpha_i)~,
\nnb \\
&&\langle \gamma(q) \vert \bar q(x) g_s \tilde G_{\mu \nu}(v x)
\gamma_\alpha \gamma_5 q(0) \vert 0 \rangle = e_q f_{3 \gamma}
q_\alpha (\varepsilon_\mu^{(\gamma)} q_\nu - \varepsilon_\nu^{(\gamma)} q_\mu) \int {\cal
D}\alpha_i e^{i (\alpha_{\bar q} + v \alpha_g) (q\cdot x)} {\cal
A}(\alpha_i)~,
\nnb \\
&&\langle \gamma(q) \vert \bar q(x) g_s G_{\mu \nu}(v x) i
\gamma_\alpha q(0) \vert 0 \rangle = e_q f_{3 \gamma} q_\alpha
(\varepsilon_\mu^{(\gamma)} q_\nu - \varepsilon_\nu^{(\gamma)} q_\mu) \int {\cal
D}\alpha_i e^{i (\alpha_{\bar q} + v \alpha_g) (q\cdot x)} {\cal
V}(\alpha_i)~, \nnb \\ 
&& \langle \gamma(q) \vert \bar q(x)
\sigma_{\alpha \beta} g_s G_{\mu \nu}(v x) q(0) \vert 0 \rangle  =
e_q \qq \left\{
        \left[\left(\varepsilon_\mu^{(\gamma)} - q_\mu \frac{\varepsilon^{(\gamma)}\mcdot x}{q\mcdot x}\right)\left(g_{\alpha \nu} -
        \frac{1}{q\mcdot x} (q_\alpha x_\nu + q_\nu x_\alpha)\right) \right. \right. q_\beta
\nnb \\ && -
         \left(\varepsilon_\mu^{(\gamma)} - q_\mu \frac{\varepsilon^{(\gamma)}\mcdot x}{q\mcdot x}\right)\left(g_{\beta \nu} -
        \frac{1}{q\mcdot x} (q_\beta x_\nu + q_\nu x_\beta)\right) q_\alpha
\nnb \\ && -
         \left(\varepsilon_\nu^{(\gamma)} - q_\nu \frac{\varepsilon^{(\gamma)}\mcdot x}{q\mcdot x}\right)\left(g_{\alpha \mu} -
        \frac{1}{q\mcdot x} (q_\alpha x_\mu + q_\mu x_\alpha)\right) q_\beta
\nnb \\ &&+
         \left. \left(\varepsilon_\nu^{(\gamma)} - q_\nu \frac{\varepsilon^{(\gamma)}\mcdot x}{q.x}\right)\left( g_{\beta \mu} -
        \frac{1}{q\mcdot x} (q_\beta x_\mu + q_\mu x_\beta)\right) q_\alpha \right]
   \int {\cal D}\alpha_i e^{i (\alpha_{\bar q} + v \alpha_g) (q\cdot x)} {\cal T}_1(\alpha_i)
\nnb \\ &&+
        \left[\left(\varepsilon_\alpha^{(\gamma)} - q_\alpha \frac{\varepsilon^{(\gamma)}\mcdot x}{q\mcdot x}\right)
        \left(g_{\mu \beta} - \frac{1}{q\mcdot x}(q_\mu x_\beta + q_\beta x_\mu)\right) \right. q_\nu
\nnb \\ &&-
         \left(\varepsilon_\alpha^{(\gamma)} - q_\alpha \frac{\varepsilon^{(\gamma)}\mcdot x}{q\mcdot x}\right)
        \left(g_{\nu \beta} - \frac{1}{q\mcdot x}(q_\nu x_\beta + q_\beta x_\nu)\right)  q_\mu
\nnb \\ && -
         \left(\varepsilon_\beta^{(\gamma)} - q_\beta \frac{\varepsilon^{(\gamma)}\mcdot x}{q\mcdot x}\right)
        \left(g_{\mu \alpha} - \frac{1}{q\mcdot x}(q_\mu x_\alpha + q_\alpha x_\mu)\right) q_\nu
\nnb \\ &&+
         \left. \left(\varepsilon_\beta^{(\gamma)} - q_\beta \frac{\varepsilon^{(\gamma)}\mcdot x}{q\mcdot x}\right)
        \left(g_{\nu \alpha} - \frac{1}{q\mcdot x}(q_\nu x_\alpha + q_\alpha x_\nu) \right) q_\mu
        \right]
    \int {\cal D} \alpha_i e^{i (\alpha_{\bar q} + v \alpha_g) (q\cdot x)} {\cal T}_2(\alpha_i)
\nnb \\ &&+
        \frac{1}{q\mcdot x} (q_\mu x_\nu - q_\nu x_\mu)
        (\varepsilon_\alpha^{(\gamma)} q_\beta -
\varepsilon_\beta^{(\gamma)} q_\alpha)
    \int {\cal D} \alpha_i e^{i (\alpha_{\bar q} + v \alpha_g) (q\cdot x)} {\cal T}_3(\alpha_i)
\nnb \\ &&+
        \left. \frac{1}{q\mcdot x} (q_\alpha x_\beta - q_\beta x_\alpha)
        (\varepsilon_\mu^{(\gamma)} q_\nu - \varepsilon_\nu^{(\gamma)} q_\mu)
    \int {\cal D} \alpha_i e^{i (\alpha_{\bar q} + v \alpha_g) (q\cdot x)} {\cal T}_4(\alpha_i)
                        \right\}~, \nnb \\
&&\langle \gamma(q) \vert \bar q(x) e_q F_{\mu\nu} (vx) q(0) \vert 0 \rangle =
-i e_q \qq \left(\varepsilon_\mu^{(\gamma)} q_\nu -
- \varepsilon_\nu^{(\gamma)} q_\mu \right)
\int {\cal D}\alpha_i e^{i (\alpha_{\bar q} + v \alpha_g) (q\cdot x)}
{\cal S}^\gamma (\alpha_i)~,\nnb \\
&&\langle \gamma(q) \vert \bar q(x) \sigma_{\alpha\beta} F_{\mu\nu}
(vx) q(0) \vert 0 \rangle = e_q \qq {1\over q\mcdot x} (q_\alpha x_\beta - q_\beta x_\alpha)
\left(\varepsilon_\mu^{(\gamma)} q_\nu - \varepsilon_\nu^{(\gamma)} q_\mu \right) \nnb \\
&&\times \int {\cal D}\alpha_i e^{i (\alpha_{\bar q} + v \alpha_g) (q\cdot
x)}
{\cal T}_4^\gamma (\alpha_i)\nnb
\eea
where $\varphi_\gamma(u)$ is the leading twist--2, $\psi^v(u)$,
$\psi^a(u)$, ${\cal A}$ and ${\cal V}$ are the twist--3, and
$h_\gamma(u)$, $\mathbb{A}$, ${\cal S}$, $\widetilde{\cal S}$,
${\cal S}^\gamma$,
${\cal T}_i$ ($i=1,~2,~3,~4$), ${\cal T}_4^\gamma$ are the
twist--4 photon DAs, $\chi$ is the magnetic susceptibility, and
the measure ${\cal D} \alpha_i$ is given by,
\bea
\int {\cal D} \alpha_i = \int_0^1 d \alpha_{\bar q} \int_0^1 d
\alpha_q \int_0^1 d \alpha_g \delta(1-\alpha_{\bar
q}-\alpha_q-\alpha_g)~.\nnb
\eea
After calculating the correlation functions in the deep Euclidean domain, and
separating the coefficients of the structures
$\epsilon_{\alpha\mu\lambda\tau} p^\lambda q^\tau q^\nu$ for the $T_Q \to P_Q
\gamma $ transition, and the coefficients of the structures
$(\varepsilon^{(\gamma)} \mcdot p)q _\mu g_{\nu\rho}$,
$(\varepsilon^{(\gamma)} \mcdot p) p_\mu q_\nu q_\rho$, and
$\varepsilon_\rho^{(\gamma)} q_\mu q_\nu$ for the $T_Q \to V_Q \gamma$ transition, and
matching them with the corresponding coefficients of these structures on the
phenomenological side, we get the desired sum rules for the transition form factors at $Q^2=0$.
We then perform the Borel transformations over the variables $(-p^2)$ and $-(p+q)^2$ in order
to suppress the contributions of the high states and the continuum, and
equate the coefficients of the aforementioned structures, from which we
obtain the following sum rules for the corresponding form factors,
\bea
\label{eesr14}
- {f_{T_Q} m_{T_Q}^3 f_{P_Q} m_{P_Q}^2 \over 2(m_Q+m_q)} e^{-(m_{P_Q}^2/M_1^2 +
m_{T_Q}^2/M_2^2)} \, g \es \Pi^{(P)}~,\\ \nnb \\
\label{eesr15}
- {f_{T_Q} f_{V_Q} m_{T_Q}^3 m_{V_Q}  \over (m_{T_Q}^2-m_{V_Q}^2)
}\,e^{-(m_{V_Q}^2/M_1^2 + m_{T_Q}^2/M_2^2)} h_1 \es \Pi_1^{(V)}~, \nnb\\
{f_{T_Q} f_{V_Q} m_{T_Q} \over 4 m_{V_Q}}
\,e^{-(m_{V_Q}^2/M_1^2 + m_{T_Q}^2/M_2^2)} \Big[ 2 h_1 
+ \left(m_{T_Q}^2 + m_{V_Q}^2\right) h_2 - 4 m_{V_Q}^2 h_3 \Big]
\es \Pi_2^{(V)}~,\nnb\\
- f_{T_Q} f_{V_Q} m_{T_Q}^3 m_{V_Q}
\,e^{-(m_{V_Q}^2/M_1^2 + m_{T_Q}^2/M_2^2)} h_3 \es \Pi_3^{(V)}~. 
\eea
The expressions of the invariant functions $\Pi^{(P)}$ and  $\Pi_i^{(V)}$
are presented in the Appendix.

The continuum subtraction procedure for the LCSR is given in detail in
\cite{Resr20}. In our calculations we set $M_1^2=M_2^2=2 M^2$ (in this case
$u_0 \to 1/2$), and the subtraction is performed by using the formula,
\bea
\label{nolabel07}
\left(M^2\right)^n e^{-(m_Q^2/M^2)} \to {1\over\Gamma(n)}
\int_{m_Q^2}^{s_0} ds \,e^{-s/M^2}
\left(s-m_Q^2\right)^{(n-1)}~~(n\ge 1)~.\nnb
\eea
Few words about the choice of the  $M_1^2=M_2^2=2 M^2$ are in order. The sum
rules for the transition between the states with different masses require
two Borel mass parameters. The difference between the masses of the initial
and final states is small, hence we can take $M_1^2=M_2^2=2 M^2$. Obviously,
taking equal Borel mass parameters would cause additional uncertainty, and
we estimate that this uncertainty is about $(10-15)\%$.

We see from Eq. (\ref{eesr14}) that the form factor $g$ responsible for
the $T_Q \to V_Q \gamma$ decay can directly be calculated.
In order to determine the form factors $h_i$ 
for the $T_Q \to P_Q \gamma$ decay the coupled set of equations given in 
Eq. (\ref{eesr15}) should be solved where $h_i$ are expressed in term of the
combinations of the $\Pi_i^{(B)}$.

\section{Numerical analysis}

In this section, we shall perform a numerical analysis of the sum rules
obtained in the previous section, for the
relevant transition form factors at the point $Q^2=0$.

\begin{table}[h]

\renewcommand{\arraystretch}{1.3}
\addtolength{\arraycolsep}{-0.5pt}
\small
$$
\begin{array}{|l|c|} 
\hline \hline  

\qq(1~GeV)             &   (-0.246^{+0.028}_{-0.019})^3~GeV^3~\cite{Resr24}   \\
\sp(1~GeV)             &    0.8\times(-0.246^{+0.028}_{-0.019})^3~GeV^3~\cite{Resr25}  \\
m_0^2                  &   (0.8   \pm 0.1)~GeV^2~\cite{Resr25} \\
m_s(2~GeV)             &   96^{+8}_{-4}~MeV~\cite{Resr06}\\ 
f_{{\cal D}_2}         &   0.0228 \pm 0.0068~\cite{Resr26} \\
f_{{{\cal D}_{s_2}}}   &   0.023  \pm 0.011~\cite{Resr27} \\
f_{B_2}                &   0.0050 \pm 0.0005~\cite{Resr28} \\
f_{{B_{s_2}}}          &   0.0060 \pm 0.0005~\cite{Resr28} \\
f_{\cal D}             &   (0.210 \pm 0.011)~GeV~\cite{Resr28} \\
f_{{\cal D}_s}         &   (0.259 \pm 0.010)~GeV~\cite{Resr28} \\
f_{{\cal D}^\ast}      &   (0.263 \pm 0.021)~GeV~\cite{Resr28} \\    
f_{{\cal D}_s^\ast}    &   (0.308 \pm 0.021)~GeV~\cite{Resr28} \\
f_{B}                  &   (0.192 \pm 0.013)~GeV~\cite{Resr28} \\
f_{B_s}                &   (0.231 \pm 0.016)~GeV~\cite{Resr28} \\    
f_{B^\ast}             &   \left(0.196^{+0.028}_{-0.027}\right)~GeV~\cite{Resr28} \\   
f_{B_s^\ast}           &   (0.255 \pm 0.019)~GeV~\cite{Resr28} \\
\hline \hline
  
\end{array}
$$
\caption{The values of the input parameters.}
\renewcommand{\arraystretch}{1}
\addtolength{\arraycolsep}{-1.0pt}

\end{table}

The magnetic susceptibility $\chi$ is estimated within the LCSR
in \cite{Resr21} to have the value 
$\chi(1~GeV)= -(2.85 \pm 0.50)~GeV^{-2}$, which we shall use in further
numerical calculations. For the heavy quark masses we have used their
$\overline{MS}$ values, i.e., $\bar{m}_c(\bar{m}_c)= (1.28  \pm 0.003)~GeV$,
$\bar{m}_b(\bar{m}_b)=(4.16 \pm 0.03)~GeV$ \cite{Resr22,Resr23}. 
The values of the other input parameters are presented in Table 1.
In addition to these
input parameters, the LCSR also contain the main non--perturbative
parameters, namely, DAs. The expression of the photon DAs \cite{Resr17}
which we need in our calculations are:
\bea
\label{eesr16}
\varphi_\gamma(u) \es 6 u \bar u \Big[ 1 + \varphi_2(\mu)
C_2^{\frac{3}{2}}(u - \bar u) \Big]~,
\nnb \\
\psi^v(u) \es 3 [3 (2 u - 1)^2 -1 ]+\frac{3}{64} (15
w^V_\gamma - 5 w^A_\gamma)
                        [3 - 30 (2 u - 1)^2 + 35 (2 u -1)^4]~,
\nnb \\
\psi^a(u) \es [1- (2 u -1)^2] [ 5 (2 u -1)^2 -1 ]
\frac{5}{2}
    \Bigg(1 + \frac{9}{16} w^V_\gamma - \frac{3}{16} w^A_\gamma
    \Bigg)~,
\nnb \\
{\cal A}(\alpha_i) \es 360 \alpha_q \alpha_{\bar q} \alpha_g^2
        \Bigg[ 1 + w^A_\gamma \frac{1}{2} (7 \alpha_g - 3)\Bigg]~,
\nnb \\
{\cal V}(\alpha_i) \es 540 w^V_\gamma (\alpha_q - \alpha_{\bar q})
\alpha_q \alpha_{\bar q}
                \alpha_g^2~,
\nnb \\
h_\gamma(u) \es - 10 (1 + 2 \kappa^+ ) C_2^{\frac{1}{2}}(u
- \bar u)~,
\nnb \\
\mathbb{A}(u) \es 40 u^2 \bar u^2 (3 \kappa - \kappa^+ +1 ) +
        8 (\zeta_2^+ - 3 \zeta_2) [u \bar u (2 + 13 u \bar u) + 
                2 u^3 (10 -15 u + 6 u^2) \ln(u) \nnb \\ 
\ar 2 \bar u^3 (10 - 15 \bar u + 6 \bar u^2)
        \ln(\bar u) ]~,
\nnb \\
{\cal T}_1(\alpha_i) \es -120 (3 \zeta_2 + \zeta_2^+)(\alpha_{\bar
q} - \alpha_q)
        \alpha_{\bar q} \alpha_q \alpha_g~,
\nnb \\
{\cal T}_2(\alpha_i) \es 30 \alpha_g^2 (\alpha_{\bar q} - \alpha_q)
    [(\kappa - \kappa^+) + (\zeta_1 - \zeta_1^+)(1 - 2\alpha_g) +
    \zeta_2 (3 - 4 \alpha_g)]~,
\nnb \\
{\cal T}_3(\alpha_i) \es - 120 (3 \zeta_2 - \zeta_2^+)(\alpha_{\bar
q} -\alpha_q)
        \alpha_{\bar q} \alpha_q \alpha_g~,
\nnb \\
{\cal T}_4(\alpha_i) \es 30 \alpha_g^2 (\alpha_{\bar q} - \alpha_q)
    [(\kappa + \kappa^+) + (\zeta_1 + \zeta_1^+)(1 - 2\alpha_g) +
    \zeta_2 (3 - 4 \alpha_g)]~,\nnb \\
{\cal S}(\alpha_i) \es 30\alpha_g^2\{(\kappa +
\kappa^+)(1-\alpha_g)+(\zeta_1 + \zeta_1^+)(1 - \alpha_g)(1 -
2\alpha_g)\nnb \\ 
\ar\zeta_2
[3 (\alpha_{\bar q} - \alpha_q)^2-\alpha_g(1 - \alpha_g)]\}~,\nnb \\
\tilde {\cal S}(\alpha_i) \es-30\alpha_g^2\{(\kappa -
\kappa^+)(1-\alpha_g)+(\zeta_1 - \zeta_1^+)(1 - \alpha_g)(1 -
2\alpha_g)\nnb \\ 
\ar\zeta_2 [3 (\alpha_{\bar q} -
\alpha_q)^2-\alpha_g(1 - \alpha_g)]\}.
\eea
The constants entering  the above DAs are borrowed from
\cite{Resr17,Resr29} whose values are given in Table (2).

\begin{table}[h]

\renewcommand{\arraystretch}{1.3}
\addtolength{\arraycolsep}{-0.5pt}
\small
$$
\begin{array}{|c|c|c|c|c|c|c|c|c|c|}
\hline \hline
\varphi_2  & \kappa  &  \kappa^+ & \xi_1  & \xi_1^+  & \xi_2 
& \xi_2^+  & f_{3\gamma}~(GeV^2) & \omega_\gamma^V  & \omega_\gamma^A \\ \hline 
0.0  & 0.2     &   0.0     & 0.4    & 0.0      &  0.3   &  0.0 
&  (-4.0 \pm 2.0 )\times10^{-3} & 3.8 \pm 1.8  & -2.1 \pm 1.0 \\
\hline \hline

\end{array}
$$
\caption{The values of the constant parameters entering into
the distribution amplitudes at the renormalization scale $\mu=1~GeV$
\cite{Resr17,Resr29}.}
\renewcommand{\arraystretch}{1}
\addtolength{\arraycolsep}{-1.0pt}

\end{table}

Besides the input parameters that are presented in Tables (1) and (2), sum
rules contain two more extra parameters, namely, the continuum threshold
$s_0$ and the Borel mass parameter $M^2$. The sum rules calculations demand
that the physical calculations should not depend on these auxiliary
parameters. Hence the working region of $M^2$ is determined by requiring
that the following conditions are satisfied:
i) Suppression of the  continuum and higher states contributions
compared to the pole contribution; ii) the dominance of the
perturbative contributions over the non--perturbative ones, and iii)
convergence of the OPE series. The upper bound of $M^2$ is determined by the
condition that, the higher states contribution should be less than $40\%$
with respect to the contributions coming from the perturbative ones, i.e.,
\bea
\label{nolabel08}
{\int_{m_Q^2}^{s_0} \rho (s) e^{-s/M^2} \, ds \over
 \int_{m_Q^2}^{\infty}  \rho (s) e^{-s/M^2} \, ds} < 0.4~. \nnb 
\eea
The lower limit of $M^2$ is obtained from the condition that the OPE series,
that is, the higher twist contributions should be smaller compared to the
leading twist contributions. These conditions lead to the following working
region of $M^2$
\bea
\label{nolabel09}
&&\phantom{1}4.0~GeV^2  \le M^2 \le 10.0~GeV^2~\mbox{for the}~{\cal D}_2 \to 
{\cal D}({\cal D}^\ast) \gamma~, \nnb \\
&&12.0~GeV^2 \le M^2 \le 20.0~GeV^2~\mbox{for the}~B_2 \to B(B^\ast) \gamma~.\nnb
\eea
 
The continuum threshold $s_0$ is obtained from the analysis of the mass
sum rules, and are given as:
$(s_0)_{{\cal D}_2} = (8.5 \pm 0.5)~GeV^2$, $(s_0)_{{\cal D}_{s_2}} = (9.5 \pm
0.5)~GeV^2$ \cite{Resr16,Resr27}, $(s_0)_{B_2} = (39 \pm 1)~GeV^2$,
$(s_0)_{B_{s_2}} = (41 \pm 1)~GeV^2$ \cite{Resr28}.

Having determined the working regions of $M^2$ and $s_0$, we now study the
dependence of $g$ and $h_i$ on $M^2$, at several fixed values of $s_0$.
We observe that indeed $g$ and $h_i$ demonstrate good stability with respect
to the variation in $M^2$ in its working region.

The dependencies of $g$ and $h_i$ on $s_0$ at several fixed values of $M^2$
are also analyzed. We find that these couplings exhibit very weak dependence
on the variation of $s_0$.
Our final results for the transition form factors 
$g$ and $h_i$ are presented in Table 3.
\begin{table}[h]

\renewcommand{\arraystretch}{1.3}
\addtolength{\arraycolsep}{-0.5pt}
\small
$$
\begin{array}{|l|c|c|c|c|}
\hline \hline
   & g~(GeV^{-2})  & h_1  &  h_2~(GeV^{-2}) & h_3~(GeV^{-2}) \\ \hline
{\cal D}_2^0 \to {\cal D}^0 \gamma                 & -0.34  \pm 0.03  &                   &                 &                \\
{\cal D}_2^+ \to {\cal D}^+ \gamma                 &  0.26  \pm 0.03  &  &   &\\
{\cal D}_{s_2} \to {\cal D}_s \gamma            &  0.13  \pm 0.03  &   &   &\\
{\cal D}_2^0 \to {\cal D}^{\ast 0} \gamma            &                  & -0.025 \pm 0.001  &  1.21 \pm 0.19  &  0.87 \pm 0.13 \\
{\cal D}_2^+ \to {\cal D}^{\ast +}\gamma            &                  & -0.15 \pm 0.02    & -1.40 \pm 0.30  & -0.90 \pm 0.20 \\
{\cal D}_{s_2} \to {\cal D}_s^\ast \gamma       &                  & -0.13 \pm 0.02    & -1.10 \pm 0.20  & -0.70 \pm 0.08 \\
B_2^- \to B^- \gamma                               & -0.35  \pm 0.03  &                   &                 &                \\
B_2^0 \to B^0 \gamma                               &  0.14  \pm 0.02  &                   &                 &                \\
B_{s_2} \to B_s \gamma                          &  0.080 \pm0.004  &                   &                 &                \\
B_2^- \to B^{\ast -} \gamma                          &                  & 0.95 \pm 0.10  &  4.60 \pm 0.60  &  2.80 \pm 0.40    \\
B_2^0 \to B^{\ast 0}\gamma                          &                  & 0.17 \pm 0.03  & -1.30 \pm 0.15  & -1.10 \pm 0.11    \\
B_{s_2} \to B_s^\ast \gamma                     &                  & 0.12 \pm 0.02  & -1.00 \pm 0.10  & -0.80 \pm 0.04    \\
\hline \hline

\end{array}
$$
\caption{The values of the form factors $g$ and
$h_i$, at the the point $Q^2=0$.}
\renewcommand{\arraystretch}{1}
\addtolength{\arraycolsep}{-1.0pt}

\end{table}

Having obtained the form factors, the decay widths of the
corresponding transitions can be estimated. The width(s) for the generic
$A \to B \gamma$ is given by the following expression:
\bea
\label{nolabel13}
\Gamma(A \to B \gamma) = {\alpha \over 4 m_A^3} {1\over 2 s_A + 1}
(m_A^2 - m_B^2) \vel M \ver^2~, \nnb 
\eea
where $s_A$ is the spin of the initial particle $A$.
Using this expression for the decay width and substituting the values of $g$
and $h_i$ from Table 3, below we list the numerical values for the
decay widths of the radiative decays the heavy--light tensor mesons under
consideration:
\bea
\label{nolabel12}
\Gamma({\cal D}_2^0    \to {\cal D}^0 \gamma)         \es ( 3.19 \pm 0.54)~KeV~, \nnb \\
\Gamma({\cal D}_2^+    \to {\cal D}^+ \gamma)         \es ( 1.86 \pm 0.46)~KeV~, \nnb \\
\Gamma({\cal D}_{s_2} \to {\cal D}_s \gamma)       \es ( 2.32 \pm 0.46)~KeV~, \nnb \\
\Gamma({\cal D}_2^0    \to {\cal D}^{\ast 0}\gamma)    \es ( 5.54 \pm 1.69)~KeV~, \nnb \\
\Gamma({\cal D}_2^+    \to {\cal D}^{\ast +}\gamma)    \es (15.40 \pm 5.10)~KeV~, \nnb \\
\Gamma({\cal D}_{s_2} \to {\cal D}_s^\ast \gamma)) \es ( 10.20 \pm 3.50)~KeV~, \nnb \\
\Gamma(B_2^- \to B^- \gamma)                          \es ( 1.61  \pm 0.29)~KeV~, \nnb \\
\Gamma(B_2^0 \to B^0 \gamma)                          \es ( 0.26  \pm 0.08)~KeV~, \nnb \\
\Gamma(B_{s_2} \to B_s \gamma)                     \es ( 0.088 \pm 0.008)~KeV~, \nnb \\
\Gamma(B_2^- \to B^{\ast -}\gamma)                     \es (50.70  \pm 11.22)~KeV~, \nnb \\
\Gamma(B_2^0 \to B^{\ast 0}\gamma)                     \es ( 0.60  \pm 0.17)~KeV~, \nnb \\
\Gamma(B_{s_2} \to B_s^\ast \gamma))               \es ( 0.36  \pm 0.08)~KeV~. \nnb
\eea
Using the experimental values of the decay widths of the tensor mesons under
consideration, which are given as: $\Gamma({\cal D}_2) = (46.7 \pm
1.2)~MeV$, $\Gamma({\cal D}_{s_2}) = (16.9 \pm 0.8)~MeV$,
$\Gamma(B_2) = (20 \pm 5)~MeV$, and $\Gamma(B_{s_2}) = (1.47 \pm 0.33)~MeV$,
we observe that the branching ratios of the considered radiative decays are of the
order of $10^{-3} \div 10^{-5}$. Referring to these results we can comment
that the radiative decays of the heavy--light tensor mesons are quite
accessible at LHCb.

\section{Conclusion}

In the present work the radiative decays of the tensor
mesons to heavy--light pseudoscalar and vector mesons are studied
within the LCSR. For this
purpose, first the transition form factors entering into the matrix
element of the relevant decays are calculated. Using the values of the
relevant form factors at the point $Q^2=0$
the corresponding decay widths are estimated. It is observed that
the branching ratios of the considered decays are larger than $10^{-5}$, and
therefore they could potentially be measured at LHCb in the near future.


\newpage


\section*{Appendix: Expressions of the invariant functions $\Pi^{(P)}$ and
$\Pi_i^{(V)}$}


In this Appendix we present the expressions of the invariant functions
$\Pi^{(P)}$ and $\Pi_i^{(V)}$ in Eqs. (\ref{eesr14}) and (\ref{eesr15}).


\section*{$T_Q \to P_Q \gamma$ transition}
{\bf $\epsilon_{\alpha\mu\lambda\tau} p^\lambda q^\tau q^\nu$ structure:}
\bea
\Pi^{(P)} \es
{e^{-m_Q^2/M^2}\over 6912 M^8}
e_q \GG m_Q^4 \qq (1 + 2 u_0) \mathbb{A}(u_0) \nnb \\
\ar {e^{-m_Q^2/M^2}\over 6912 M^6 \pi^2}
m_Q^2 \Big\{48 e_Q m_0^2 m_Q m_q \pi^2 \qq + 
   e_q \GG \Big[3 m_q e^{m_Q^2/M^2} ({\cal J}_0 + m_Q^2 {\cal J}_1) \nnb \\
\ar 2 \pi^2 \qq(1 - 2 u_0) \mathbb{A}(u_0) + 2 \pi^2 f_{3\gamma} m_Q (1 + 2 u_0) 
        \psi^a(u_0)\Big]\Big\} \nnb \\
\ar {e^{-m_Q^2/M^2}\over 6912 M^4 \pi^2}
m_Q \Big\{144 e_Q m_0^2 m_Q \pi^2 \qq - 3 e_q \GG m_Q m_q \Big[2 \gamma_E
- e^{m_Q^2/M^2} \Big(2 {\cal I}_1 + {\cal J}_1 \nnb \\
\ar 2 m_Q^2 ({\cal I}_2 + {\cal J}_2)\Big)\Big] - 
   2 e_q \GG \pi^2 \Big[2 m_Q \qq (1 + 2 u_0) \chi \varphi_\gamma(u_0) - 
     f_{3\gamma} (1 - 6 u_0) \psi^a(u_0)\Big]\Big\} \nnb \\
\ek {e^{-m_Q^2/M^2}\over 2304 M^2 \pi^2}
\Big\{ e_q \GG [2 m_Q + m_q (1 + 2 \gamma_E)] - 48 e_Q (m_0^2 - 2 m_Q m_q) 
    \pi^2 \qq \nnb \\
\ar e_q m_Q^2 \Big[24 \pi^2 \qq (1 + 2 u_0) \mathbb{A}(u_0) - 
     \GG m_q e^{m_Q^2/M^2} \Big(2 {\cal I}_2 + {\cal J}_2 + m_Q^2 (3 {\cal I}_3 + 
2 {\cal J}_3)\Big)\Big]\Big\} \nnb \\
\ek {e^{-m_Q^2/M^2}\over 1728 m_Q M^2}
e_q \GG \Big[2 m_Q \qq \chi \varphi_\gamma(u_0) + 
f_{3\gamma} \psi^a(u_0)\Big] \nnb \\
\ek {M^2\over 32 \pi^2}
m_Q^3 \Big\{2 e_Q [{\cal I}_2 - m_Q (m_Q + m_q) {\cal I}_3] - 
     e_q m_Q \Big[2 m_Q {\cal I}_3 - 2 m_q {\cal J}_3 - 
       m_Q^2 m_q ({\cal I}_4 + 3 {\cal J}_4)\Big]\Big\} \nnb \\
\ar {e^{-m_Q^2/M^2}\over 24} M^2 
e_q \qq (1 + 2 u_0) \chi \varphi_\gamma(u_0)~. \nnb
\eea


\section*{$T_Q \to V_Q \gamma$ transition}
{\bf $(\varepsilon^{(\gamma)} \mcdot p)q _\mu g_{\nu\rho}$ structure:}
%
\bea
\Pi_1^{(V)} \es 
- {M^4 \over 16 \pi^2} (e_Q - e_q) m_Q^6 (3 {\cal I}_4 - 4 m_Q^2 {\cal I}_5) \nnb \\
\ar {M^2 \over 16 \pi^2} 
m_Q^4 \Big\{2 e_q m_Q m_q {\cal I}_3 + e_Q \Big[{\cal I}_2 - m_Q (m_Q +
2 m_q) {\cal I}_3\Big]\Big\} \nnb \\
\ek {M^2 e^{-m_Q^2/M^2} \over 24}  \Big\{e_q f_{3\gamma} \Big[4
\widetilde{j}_1(\psi^v)  +
\psi^a(u_0)\Big]\Big\} \nnb \\
\ek {e^{-m_Q^2/M^2} \over 432 M^6} m_Q^3 \qq \Big[6 e_Q m_0^2 m_Q m_q - e_q \GG
\widetilde{j}_2(h_\gamma)\Big] \nnb \\
\ek {e^{-m_Q^2/M^2} \over 1728 M^4}  m_Q \Big\{24 e_Q m_0^2 m_Q (3 m_Q - m_q) \qq + 
e_q \GG \Big[4 \qq \widetilde{j}_2(h_\gamma) \nnb \\
\ek f_{3\gamma} m_Q \Big(4
\widetilde{j}_1(\psi^v)  + \psi^a(u_0)\Big)\Big]\Big\} \nnb \\
\ek {e^{-m_Q^2/M^2} \over 1728 \pi^2 m_Q M^2}
\Big\{3 e_q \GG m_Q^2 m_q - 
   72 e_Q m_Q \Big[2 m_Q^2 m_q - m_0^2 (m_Q - m_q)\Big] \pi^2 \qq \nnb \\
\ar e_q \GG \pi^2 \Big[4 \qq \widetilde{j}_2(h_\gamma) - f_{3\gamma} m_Q 
      \Big(4 \widetilde{j}_1(\psi^v)  + \psi^a(u_0)\Big)\Big]\Big\} \nnb \\
\ar {e^{-m_Q^2/M^2}\over 1152  m_Q \pi^2} 
\Big[96 e_Q m_Q (2 m_Q - m_q) \pi^2 \qq +
    e_q \GG (m_Q + 2 m_q - 4 m_Q^5 e^{m_Q^2/M^2} {\cal I}_3)\Big] \nnb \\
\ek{e^{-m_Q^2/M^2}\over 6} \Big[e_q m_Q \qq \widetilde{j}_2(h_\gamma)\Big]~. \nnb
\eea


{\bf $(\varepsilon^{(\gamma)} \mcdot p) p_\mu q_\nu q_\rho$ structure:}
\bea
\Pi_2^{(V)} \es
{M^2 \over 16 \pi^2} m_Q^4 \Big[ (2 e_Q - e_q) {\cal I}_3) - 
m_Q^2 (6 e_Q - 4 e_q) {\cal I}_4 + 4 (e_Q - e_q) m_Q^4 {\cal
I}_5 \Big] \nnb \\
\ek {e^{-m_Q^2/M^2}\over 864 M^8}
e_q \GG m_Q^3 \qq \Big[(1 + 2 u_0) \widetilde{j}_2(h_\gamma) + 
     2 \widetilde{j}_3(h_\gamma)\Big] \nnb \\
\ar {e^{-m_Q^2/M^2}\over 3456 M^6}
m_Q \Big\{48 e_Q m_0^2 m_Q m_q \qq + 
   e_q \GG \Big[4 f_{3\gamma} m_Q (3 + 2 u_0) \widetilde{j}_1(\psi^v)  \nnb \\
\ek     4 (1 - 6 u_0) \qq \widetilde{j}_2(h_\gamma) + 24 \qq \widetilde{j}_3(h_\gamma) + 
     f_{3\gamma} m_Q \Big(8 \widetilde{j}_2(\psi^v)  - (1 + 2 u_0)
\psi^a(u_0)\Big)\Big]\Big\} \nnb \\
\ar {e^{-m_Q^2/M^2}\over 1728 m_Q M^4}
\Big\{72 e_Q m_0^2 m_Q m_q \qq + 
   e_q \GG \Big[12 f_{3\gamma} m_Q \widetilde{j}_1(\psi^v)  + 4 \qq \widetilde{j}_2(h_\gamma) - 
     f_{3\gamma} m_Q \psi^a(u_0)\Big]\Big\} \nnb \\
\ek {e^{-m_Q^2/M^2}\over 1152 M^2 \pi^2}
\Big\{
96 e_Q m_q \pi^2 \qq - e_q \GG \Big[1 + m_Q^2 e^{m_Q^2/M^2} ({\cal I}_2 - 
     4 m_Q^2 {\cal I}_3)\Big] \nnb \\
\ek 96 e_q m_Q \pi^2 \qq \Big[(1 + 2 u_0) \widetilde{j}_2(h_\gamma) + 
2 \widetilde{j}_3(h_\gamma) \Big] \Big\} \nnb \\
\ek {e^{-m_Q^2/M^2}\over 48}
e_q f_{3\gamma} \Big\{4 (3 + 2 u_0) \widetilde{j}_1(\psi^v)  + 
8 \widetilde{j}_2(\psi^v) - (1 + 2 u_0) \psi^a(u_0)\Big\}~.\nnb
\eea


{\bf $\varepsilon_\rho^{(\gamma)} q_\mu q_\nu$ structure:}
\bea
\Pi_3^{(V)} \es
- {M^4 e^{-m_Q^2/M^2}\over 16 \pi^2}
\Big\{2 e_q \Big[1 - m_Q^6 e^{m_Q^2/M^2} ({\cal I}_4 - m_Q^2 {\cal I}_5)\Big] +
    e_Q m_Q^2 e^{m_Q^2/M^2} \Big[{\cal I}_2 - 2 m_Q^4 ({\cal I}_4 + m_Q^2 
{\cal I}_5)\Big]\Big\} \nnb\\
\ek {e^{-m_Q^2/M^2}\over 3456 M^8}
\Big[e_q \GG m_Q^5 (1 + 2 u_0) \qq \mathbb{A}(u_0)\Big] \nnb \\
\ek {1\over 1152 M^6 \pi^2}
e_q gGgG m_Q^3 m_q \Big[{\cal J}_0 + m_Q^2 {\cal J}_1\Big] \nnb \\
\ek {e^{-m_Q^2/M^2}\over 1728 M^6}
m_Q^3 \Big\{24 e_Q m_0^2 m_Q m_q \qq - e_q \GG \Big\{6 \qq u_0 \mathbb{A}(u_0) - 
      (1 + 2 u_0) \qq \widetilde{j}_1(h_\gamma) \nnb \\
\ar 4 \qq [u_0 \widetilde{j}_2(h_\gamma) + 
        \widetilde{j}_3(h_\gamma)] - f_{3\gamma} m_Q (1 + 2 u_0) \psi^a(u_0)\Big]\Big\} \nnb \\
\ar {1\over 1152 M^4 \pi^2}
e_q \GG m_Q m_q \Big[3 {\cal J}_0 - m_Q^2 (3 {\cal I}_1 - {\cal J}_1) - 
    3 m_Q^4 ({\cal I}_2 + {\cal J}_2)\Big] \nnb \\
\ar {e^{-m_Q^2/M^2}\over 6912 M^4 \pi^2} 
m_Q \Big\{12 m_Q \Big[e_q \gamma_E \GG m_Q m_q - 8 e_Q m_0^2 (3 m_Q -
2 m_q) \pi^2 \qq\Big] \nnb \\
\ar 12 e_q \GG \pi^2 \qq (1 - 2 u_0) \mathbb{A}(u_0) - 
    4 e_q \GG \pi^2 \qq (1 - 6 u_0) \widetilde{j}_1(h_\gamma) \nnb \\
\ar e_q \GG \pi^2 \Big[8 f_{3\gamma} m_Q (1 + 2 u_0) \widetilde{j}_1(\psi^v) + 
      16 \qq (2 - 3 u_0) \widetilde{j}_2(h_\gamma) - 48 \qq
\widetilde{j}_3(h_\gamma)\Big] \nnb \\
\ar 4 e_q \GG m_Q \pi^2 \Big[2 m_Q \qq (1 + 2 u_0) \chi \varphi_\gamma(u_0)
f_{3\gamma} \Big(4 \widetilde{j}_2(\psi^v) - (1 - 5 u_0) \psi^a(u_0) \nnb \\
\ar \psi^v(u_0) + 
          2 u_0 \psi^v(u_0)\Big)\Big]
- f_{3\gamma} e_q \GG m_Q \pi^2 (1 + 2 u_0) 
       \psi^{a\prime}(u_0)\Big\} \nnb \\
%
\ar {1\over 1152 M^2 \pi^2}
e_q \GG m_Q m_q \Big\{6 {\cal I}_1 + 3 {\cal J}_1 - 
    m_Q^2 \Big[{\cal I}_2 - 2 {\cal J}_2 + 
m_Q^2 (9 {\cal I}_3 + 6 {\cal J}_3)\Big]\Big\} \nnb \\
\ar {e^{-m_Q^2/M^2}\over 3456 m_Q M^2 \pi^2}
 \Big\{3 m_Q \Big[e_q \GG m_Q (2 m_Q + m_q - 2 \gamma_E m_q) + 
     8 e_Q (5 m_0^2 + 12 m_Q^2) m_q \pi^2 \qq\Big] \nnb \\
\ar e_q \pi^2 \Big[72 m_Q^4 \qq (1 + 2 u_0) \mathbb{A}(u_0) + 
     \GG \Big(4 \qq \widetilde{j}_1(h_\gamma) - 
8 \qq \widetilde{j}_2(h_\gamma) \nnb \\
\ar       m_Q \{4 m_Q \qq (1 - 6 u_0) \chi \varphi_\gamma(u_0) + 
         f_{3\gamma} [8 \widetilde{j}_1(\psi^v) + 2 \psi^a(u_0) + 4 \psi^v(u_0) - 
           \psi^{a\prime}(u_0)]\}\Big)\Big]\Big\} \nnb \\
\ek {e^{-m_Q^2/M^2}\over 576 m_Q \pi^2}
e_q \gamma_E m_q \Big(\GG + 72 m_Q^2 M^2\Big) \nnb \\
\ek {1\over 16 \pi^2}
m_Q^3 M^2 \Big\{e_Q (m_Q + 2 m_q) {\cal I}_2 - e_Q m_Q^3 {\cal I}_3 - 
    e_q m_q \Big[{\cal J}_2 + m_Q^2 ({\cal I}_3 + 2 {\cal J}_3)\Big]\Big\} \nnb \\
\ek  {e^{-m_Q^2/M^2}\over 96} 
e_q M^2 \Big\{8 (f_{3\gamma} + 2 f_{3\gamma} u_0) \widetilde{j}_1(\psi^v) + 
    8 m_Q \qq (1 + 2 u_0) \chi \varphi_\gamma(u_0) \nnb \\
\ar 4 f_{3\gamma} \Big[4 \widetilde{j}_2(\psi^v) - (2 + u_0) \psi^a(u_0) + \psi^v(u_0) + 
      2 u_0 \psi^v(u_0)\Big] - f_{3\gamma} (1 + 2 u_0) \psi^{a\prime}(u_0) \Big\} \nnb \\
\ar {1\over 1152 \pi^2}
e_q m_Q \Big\{72 m_Q^2 m_q ({\cal J}_1 + m_Q^2 {\cal J}_2) - 
    \GG \Big[2 m_Q^3 {\cal I}_3 - m_q \Big(6 {\cal I}_2 + 3 {\cal J}_2 \nnb \\
\ar m_Q^2 [{\cal I}_3 + 2 {\cal J}_3 - m_Q^2 (11 {\cal I}_4 + 
6 {\cal J}_4)]\Big)\Big]\Big\} \nnb \\
\ar {e^{-m_Q^2/M^2}\over 1728 m_Q \pi^2}
\Big\{3 e_q \GG (m_Q - m_q) + 144 e_Q m_Q (2 m_Q - m_q) \pi^2 \qq \nnb \\
\ar   4 e_q \pi^2 \Big[18 m_Q^2 \qq \Big(\mathbb{A}(u_0) + (1 + 2 u_0) \widetilde{j}_1(h_\gamma) - 
       4 [u_0 \widetilde{j}_2(h_\gamma) + \widetilde{j}_3(h_\gamma)]\Big)
\nnb \\
\ek  \GG \qq \chi \varphi_\gamma(u_0) + 18 f_{3\gamma} m_Q^3 
(1 + 2 u_0) \psi^a(u_0)\Big]\Big\} \nnb \\
\eea
where $s_0$ is the continuum threshold, $M_1^2=M_2^2=2 M^2$, and
\bea
u_0={M_1^2 \over M_1^2 +M_2^2}={1\over 2}~,~~~~~M^2={M_1^2 M_2^2 \over M_1^2 +M_2^2}~.\nnb
\eea

The integral functions ${\cal I}_n,~(n=1,\cdots,5)$; ${\cal J}_n,~(n=0,\cdots,4)$;
$\widetilde{j}_n(f(u_0))~(n=1,2,3)$ are defined as:
\bea
{\cal I}_n \es \int_{m_b^2}^{s_0} ds\, {e^{-s/M^2} \over s^n}~,\nnb \\
{\cal J}_n \es \int_{m_b^2}^{s_0} ds\, {\ell n\left[M^2 (s - m_Q^2)/
(\Lambda^2 s)\right] \over s^n} e^{-s/M^2}~,\nnb \\
\widetilde{j}_n(f) \es \int_{u_0}^1 du (u-u_0)^{(n-1)} f(u)~,~~\mbox{and,} \nnb \\
\psi^{\prime a}(u_0) \es \left.{d \psi^a(u) \over du} \right|_{u=u_0}~. \nnb
\eea

\end{document}